\documentstyle[epsfig,longtable]{aipproc}

\def\mkn{M_{K^0}^2}
\def\l7{\lambda_7}
\def\la{\langle}
\def\rag{\rangle}

\def\pieta{$\pi^0$-$\eta,\eta'\;$}

\def\epsrat{$\epsilon^\prime/\epsilon\;$}
\def\reepsrat{$\,{\rm Re}\,(\epsilon^\prime/\epsilon)\;$}
\def\optwo{${\cal O}(p^2)\;$}
\def\opfour{${\cal O}(p^4)\;$}
\def\opsix{${\cal O}(p^6)\;$}
\def\ra{\rightarrow}

\begin{document}
\begin{flushright} \normalsize 
UK/TP 00-03  \\ 
hep-ph/0006120 \\
June 2000
\end{flushright}

\title{Additional 
Isospin-Breaking Effects in ${\bf \epsilon^\prime/\epsilon}$}

\author{S. Gardner$^{*,}$\thanks{Talk at 
the 3rd Int. Conf. on Symmetries in Subatomic 
Physics, Adelaide, Mar. 13-17, 2000 --- based on work 
performed in collaboration with G. Valencia.}}
\address{
$^*$ Department of Physics and Astronomy, 
University of Kentucky, Lexington, KY 
40506-0055 USA
}

\maketitle

\begin{abstract}
Isospin-breaking effects, in 
particular those associated with 
electroweak-penguin contributions and \pieta mixing, 
have long been known to affect the Standard Model 
prediction of \epsrat in a 
significant manner. 
We have found an 
heretofore unconsidered isospin-violating effect of 
importance; namely, 
the $u$-$d$ quark mass
difference can spawn $|\Delta I|=3/2$ components 
in the matrix elements of the gluonic penguin operators.  
Using chiral perturbation theory and the factorization approximation
for the hadronic matrix elements, we find 
within a specific model for the low-energy constants that we can 
readily accommodate an increase in \epsrat by a factor of two. 
\end{abstract}

\section*{Introduction}

The recent measurement of a non-zero value of 
$\,{\rm Re}\,(\epsilon^\prime/\epsilon)$~\cite{ktev}
establishes the
existence of 
CP violation in direct decay and thus provides 
an important first check of the mechanism of CP violation
in the Standard Model (SM). Nevertheless, the
world average which emerges is 
${\rm Re}\,(\epsilon^\prime/\epsilon)=
(19.3\pm 2.4)\cdot10^{-4}$~\cite{world}, which is larger 
than the
``central'' SM prediction of 
$7.0\cdot 10^{-4}$~\cite{burasr,buraslh} by nearly  
a factor of three. 
This compels us to scrutinize
the SM prediction in further detail: we study isospin-violating
effects arising from the $u$-$d$ quark mass difference. 

Isospin violation plays an important role 
in the analysis of
$\epsilon^\prime/\epsilon$, for the latter
is predicated by the difference of 
the imaginary to real part ratios in the 
$|\Delta I|=1/2$ and $|\Delta I|=3/2$ $K\ra \pi\pi$ amplitudes. 
The differing charges of the $u$ and 
$d$ quarks engender $|\Delta I|=3/2$ electroweak penguin contributions,
whereas \pieta mixing, driven by the $u$-$d$ quark mass
difference, modifies the relative contribution of the 
$|\Delta I|=1/2$ and $|\Delta I|=3/2$ amplitudes in a significant way.

Here we describe 
isospin-breaking effects in the matrix elements of the 
gluonic penguin operators~\cite{sggv}, such as $Q_6$. 
These operators have always been thought to induce exclusively
$|\Delta I| =1/2$ transitions, 
but this is 
true only in the limit of isospin symmetry. 
The difference in the  up and down 
quark masses effectively 
distinguishes the interaction of gluons with 
up and down quarks, 
so that the $\langle\pi\pi|Q_6|K\rangle$ matrix element
possesses a $|\Delta I|=3/2$ component as well~\cite{bpipi}. 

Let us  begin by showing
why the numerical prediction of 
\epsrat is sensitive to the presence of 
isospin violation. 
The value of \epsrat is inferred from a ratio of ratios,
namely 
\begin{equation}
{\rm Re}\,\left(\frac{\varepsilon^\prime}{\epsilon}\right) = \frac{1}{6}
\left[
\Bigg|\frac {\eta_{+-}}{\eta_{00}}
\Bigg|^2 -1
\right]\;,
\end{equation}
where
\begin{equation}
\eta_{+-} \equiv \frac{{\cal A}(K_L \ra \pi^+ \pi^-)}{{\cal A}(K_S \ra 
\pi^+ \pi^-)} \approx \epsilon + \epsilon^\prime \;\;;\;\;
\eta_{00} \equiv \frac{{\cal A}(K_L \ra \pi^0 \pi^0)}{{\cal A}(K_S \ra 
\pi^0 \pi^0)}
\approx \epsilon - 2\epsilon^\prime \;.
\end{equation}
In the isospin-perfect limit, the two 
independent 
amplitudes present in $K\ra\pi\pi$ decay are 
distinguished by the isospin of the final-state pions, namely
$A_I \equiv {\cal A}(K\ra(\pi\pi)_{I})$ with $I=0,2$.  
\epsrat can thus be written
\begin{equation}
\frac{\epsilon^\prime}{\epsilon} = 
- \frac{\omega}{\sqrt{2}|\epsilon|}\xi(1 - \Omega) \;, 
\label{eratdef}
\end{equation}
with
\begin{equation}
\omega \equiv \frac{{\rm Re} A_2}{{\rm Re}\, A_0} \quad;\quad
\xi\equiv \frac{{\rm Im}\, A_0}{ {\rm Re}\, A_0} \quad;\quad
\Omega\equiv 
\frac{{\rm Im}\, A_2}{\omega {\rm Im}\, A_0} \;. 
\end{equation}
In standard
practice, $\omega\approx 1/22$ and ${\rm Re}\,A_0$  are
taken from experiment whereas ${\rm {Im}} A_{I}$ 
is computed using the operator-product expansion~\cite{burasr,buraslh}, 
that is, via
\begin{equation}
{\cal H}_{\rm eff}(|\Delta S|=1) = 4\frac{G_{\rm F}}{\sqrt{2}} V_{us}^* 
V_{ud}^{}
\sum_{i=1}^{10} C_i(\mu) Q_i(\mu)  + {\rm h.c.}
\end{equation}
The numerical value of  \epsrat is 
 driven by the matrix elements of the 
QCD penguin operator  $Q_6$  and the electroweak penguin operator  
$Q_8$~\cite{buras93}. 
Writing  $\langle Q_i \rangle_I$  as
$\langle (\pi\pi)_I | Q_i| K \rangle \equiv B_i^{((I+1)/2)} \, 
\langle (\pi\pi)_I | Q_i | K \rangle^{\rm (vac)}$,
where ``vac'' indicates the use of the vacuum saturation approximation, 
one recovers the schematic formula~\cite{burasr}
\begin{equation}
\frac{\epsilon^\prime}{\epsilon}
= 13 \,{\rm Im} \lambda_t \left[ B_6^{(1/2)}(1 - \Omega_{\eta+\eta^\prime})
- 0.4\, B_8^{(3/2)} 
\right] \;.
\end{equation}
Using $B_6^{(1/2)}=1.0$, $B_8^{(3/2)}=0.8$, and
$\Omega_{\eta+\eta^\prime}=0.25$ 
yields the  ``central'' SM value of 
$\epsilon^\prime/\epsilon
\sim 7.0\cdot 10^{-4}$~\cite{burasr},  
roughly a factor of three smaller than the measured 
value. Larger estimates of 
$B_6^{(1/2)}$, 
and hence of \epsrat, 
exist~\cite{bertolini,hambye,bijpra}; we 
investigate sources of 
 $\Omega_{\eta+\eta^\prime}$. Note that 
under  $\Omega_{\eta+\eta^\prime}\rightarrow
- \Omega_{\eta+\eta^\prime}$, 
\reepsrat $\rightarrow$ 2.2 
${\rm Re}\,(\epsilon^\prime/\epsilon)$.
Were $|B_6^{(1/2)}| \gg |B_8^{(3/2)}|$, flipping the sign
of  $\Omega_{\eta+\eta^\prime}$ would increase
\epsrat by a factor of 1.7. 

\begin{figure}[b!] 
\centerline{\epsfig{file=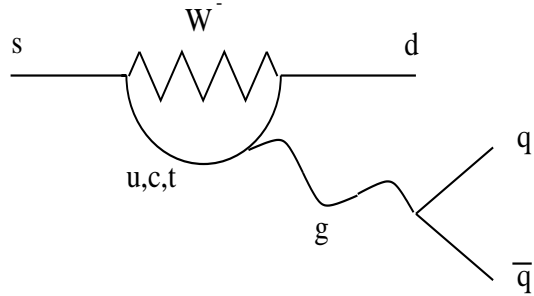,height=2.75in,width=1.5in,angle=-90}}
\vspace{5pt}
\caption{
Quark line diagram illustrating the ``strong penguin'' 
$s\ra d{\bar q} q$ transition in the Standard Model. Note
that ${\bar q} q \in {\bar u} u\,,{\bar d} d$; in the
isospin-perfect limit, $m_u=m_d$ and only $|\Delta I|=1/2$
transitions are generated. If $m_u\ne m_d$, the $K\ra\pi\pi$
matrix element associated with this operator contains
a $|\Delta I|=3/2$ component as well. }
\label{fig1}
\end{figure}

Let us consider possible sources of 
$\Omega_{\eta+\eta^\prime}$. We replace
$\Omega_{\eta+\eta^\prime}$ by $\Omega_{\rm IB}$, where
\begin{equation}
\Omega_{\rm IB} =\biggl({\sqrt{2} \over 3 \omega}\biggr) 
{{\rm Im} (A_{\rm P}(K^0 \rightarrow
\pi^+ \pi^-)-A_{\rm P}(K^0 \rightarrow\pi^0 \pi^0))\over 
{\rm Im}\,A_{\rm P}(K^0 \rightarrow \pi\pi)} 
\end{equation}
and ${\rm Im}\,A_{\rm P}(K^0 \rightarrow \pi\pi) = 
({\rm Im}\,A_{\rm P}(K^0 \rightarrow \pi^+\pi^-) + 
{\rm Im}\,A_{\rm P}(K^0 \rightarrow \pi^0\pi^0))/2$. 
``$A_{\rm P}$'' denotes an amplitude 
induced by  $(8_L,1_R) \, ({\rm e.g.}, Q_6)$   
operators --- the empirical $|\Delta I|=1/2$ rule 
suggests such operators dominate 
the isospin-violating effects.
$\Omega_{\rm IB}$  vanishes 
in the absence of isospin violation, i.e., if  $m_u=m_d$,  
$e_u=e_d$. It can be generated by both strong-interaction and 
electromagnetic effects, mediated by $m_d\ne m_u$ and 
$e_u\ne e_d$~\cite{cirigli}, respectively. We focus on 
$m_d\ne m_u$ effects. The latter include \pieta 
mixing~\cite{bijwi,dght,buge}; 
in  ${\cal O}(p^2,1/N_c)$  this yields
 $\Omega_{\eta+\eta^\prime}=0.25 \pm 0.05$~\cite{dght,buge},
used in the analysis of Ref.~\cite{burasr}. However, 
$m_u \ne m_d$  effects  can also spawn a $|\Delta I|=3/2$  component 
in the matrix elements of 
the gluonic penguin operators~\cite{bpipi}, as 
illustrated in Fig.~\ref{fig1}. We turn to a chiral
Lagrangian analysis in order to estimate the size of this effect~\cite{sggv}. 

\section*{Chiral Lagrangian Analysis}
\begin{figure}[b!] 
\centerline{\epsfig{file=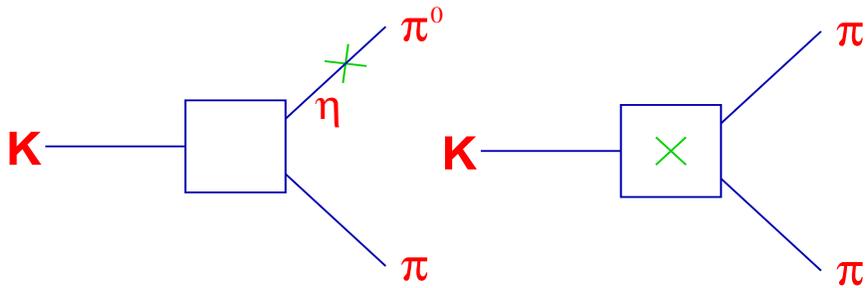,height=4.5in,width=1.5in,angle=-90}}
\vspace{5pt}
\caption{Isospin violation in $K\ra\pi\pi$ decays. The
square box represents the weak $|\Delta S|=1$ transition
at low energies, whereas the ``$\times$'' represents
the presence of $m_d\ne m_u$ effects.
Mass effects do not occur
in the weak transition in leading order 
in chiral perturbation theory, so that 
only the left-hand diagram occurs 
in \optwo and in ${\cal O}(m_d -m_u)$
--- $\pi^0$-$\eta$
mixing, mediated by $m_d - m_u$ effects in 
the strong chiral
Lagrangian, can occur. In \opfour 
both diagrams are present; the 
new effect we discuss is associated with the right-hand diagram.
}
\label{fig2}
\end{figure}

The weak chiral Lagrangian for $K\ra\pi\pi$ decay
is written in terms of the unitary matrix
 $U=\exp(i\phi/f)$  and the function $\chi$, both of
which transform as  $U \rightarrow R U L^\dagger$  
under  the chiral group $SU(3)_L\times SU(3)_R$. 
The function
 $\phi$  represents the octet of pseudo-Goldstone 
bosons, i.e., $\phi= \sum_{a=1,\dots, 8} \lambda_a \phi_a$.
In the absence of external fields, 
 $\chi = 2B_0 M$  with
 $M={\rm diag}(m_u,m_d,m_s)$  
and  $B_0 \propto \langle\bar{q}q\rangle$. 
The leading-order, ${\cal O}(p^2)$, 
weak chiral Lagrangian contains no mass-dependent
terms~\cite{cronin}, so that 
 $m_d\ne m_u$  effects 
in the hadronization of the gluonic penguin
operators first appear in ${\cal O}(p^4)$. This is illustrated in
Fig.~\ref{fig2}.

Let us enumerate the
possible isospin-violating effects which occur in 
 ${\cal O}(m_d - m_u)$  and \opfour:
\begin{itemize}
\item[i)] 
$\pi^0$-$\eta$ mixing realized from the \optwo strong
chiral Lagrangian, in concert with the \optwo weak chiral Lagrangian, 
computed to one-loop order. 

\item[ii)] 
$\pi^0$-$\eta$ mixing, 
realized from the \optwo strong chiral Lagrangian, 
combined with the isospin-conserving 
vertices of the \opfour weak chiral Lagrangian.

\item[iii)] 
$\pi^0$-$\eta$ mixing as realized from the
strong chiral Lagrangian in ${\cal O}(p^4)$,
combined with the \optwo weak chiral Lagrangian. 
The $\pi^0$-$\eta^\prime$ 
mixing effects included in Refs.~\cite{dght,buge}
ape this effect. 
 
\item[iv)] Isospin violation in the vertices of 
the \opfour weak 
chiral Lagrangian.
This serves as our focus here, for 
it contains the qualitatively new effects we argue.

\end{itemize}

We use the octet terms in the ${\cal O}(p^4)$,
CP-odd weak chiral Lagrangian of Ref.~\cite{kambor}.
Collecting the $\chi$-dependent terms as per iv),
working to  ${\cal O}(m_d-m_u)$, and dropping terms suppressed 
by $M_\pi^2/M_K^2$, we find 
\begin{equation}
\Omega_{\rm P} = {2\sqrt{2}\over 3\omega}{\mkn\over \mkn-M_\pi^2}
{B_0(m_d-m_u)\over c_2^-} {\tilde E}^- 
\approx {0.12 {\rm GeV}^{2}\over c_2^-}  {\tilde E}^- 
\label{omegaP}
\end{equation}
with  ${\tilde E}^-=  2E_1^- -2E_3^- - 4E_4^- -
E_{10}^- -E_{11}^- -4E_{12}^- -E_{15}^-$. Note that $c_2^-$ 
is the low-energy constant associated with the ${\cal O}(p^2)$,
$(8_L,1_R)$ weak chiral Lagrangian~\cite{kambor}.
As per ii), 
$\pi^0$-$\eta$ mixing in \optwo also enters when combined with 
the isospin-conserving vertices of the \opfour weak chiral
Lagrangian. Including the $\chi$-dependent octet terms, we 
find
\begin{equation}
\Omega_{\eta + \eta^\prime}^{(4)} = 
{2\sqrt{2}\over 3\omega}{\mkn\over \mkn-M_\pi^2}
{B_0(m_d-m_u)\over c_2^-} E_{\eta+\eta^\prime}^- 
\approx {0.12 {\rm GeV}^{2}\over c_2^-} E_{\eta+\eta^\prime}^-
\label{omegaetap}
\end{equation}
with $E_{\eta + \eta^\prime}^-=
- 2(E_3^- +E_4^- -E_5^-)+ (E_{10}^- -E_{11}^-)/2 
-2E_{12}^- +E_{14}^- +3E_{15}^-/2$ --- so that no manifest
cancellation with the terms of Eq.~(\ref{omegaP}) occurs.  
\begin{figure}[b!] 
\centerline{\epsfig{file=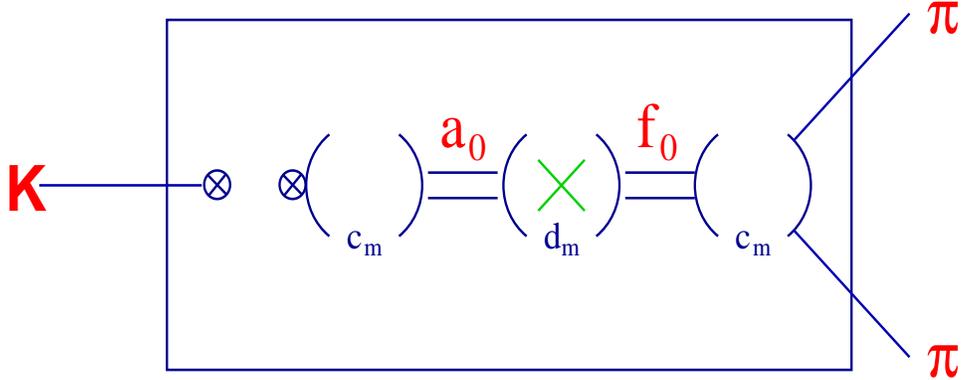,height=2.in,width=5in}}
\vspace{5pt}
\caption{
A contribution to the right-hand diagram of Fig.~\protect{\ref{fig2}}
in ${\cal O}(m_d-m_u)$, estimated in the factorization approximation 
with explicit, scalar-resonance degrees of freedom. 
The ``$\otimes$'' represents a bosonized current; the
open parentheses denote contributions from the vertices of 
the \opsix model Lagrangian of Ref.~[20].
In this case the isospin-violating
contribution is driven by $a_0$-$f_0$ mixing.}
\label{fig3}
\end{figure}

The low-energy constants  $E_i^-$   are unknown, so that
we turn to the factorization approximation to proceed. 
The construction relevant to $(8_L,1_R)$ transitions in
$K^0 \ra\pi\pi$  decay is~\cite{sekhar}
\begin{eqnarray}
{\cal L}_{\rm P} =&-{G_F\over \sqrt{2}}V_{us}^\star V_{ud} \, C_6 \,
\biggl( -8 (\bar{s}_Lq_R)(\bar{q}_R d_L)\biggr) ~+~{\rm h.c.}   \\
\rightarrow &{G_F\over \sqrt{2}}V_{us}^\star V_{ud}\, C_6 \,
32 B_0^2 {\delta {\cal L}_{\rm str}\over \delta \chi_{3i}^\dagger}
{\delta {\cal L}_{\rm str}\over \delta \chi_{i2}}~+~{\rm h.c.}  \;,  
\end{eqnarray}
where ${\cal L}_{\rm str}$  is the strong chiral Lagrangian.
To generate terms of \opfour in ${\cal L}_{\rm P}$ requires
terms of both 
${\cal O}(p^4)$~\cite{gl} and ${\cal O}(p^6)$~\cite{fearing} in 
${\cal L}_{\rm str}$.
Unfortunately, the low-energy constants of the 
latter are also unknown; 
the use of ``resonance saturation'' 
allows us to estimate 
some of them.
We explicitly
consider the scalar nonet of resonances as per Ref.~\cite{bijnens}.
An example of the manner
in which the scalar resonances can generate contributions to the
$E_i^-$ is illustrated in Fig.~\ref{fig3}.
Integrating out the scalar resonances for $p^2 \ll M_{S}^2$,
we find two terms which contribute to the scalar densities in the 
bosonization of $Q_6$~\cite{bijnens},
\begin{equation}
{\cal L}_S^{(6)}= {d_m c_m^2\over 2 M_S^4}\la\chi_+^3\rag
+{c_d c_m d_m \over M_S^4}\la\chi_+^2 L^2\rag\;,
\end{equation}
yielding contributions to $E_1^-$ and $E_{10}^-$ in terms 
of $d_m$, $c_m$, $c_d$, and $M_S$. The parameter 
$d_m$ is ill-known; we find
$d_m \sim -2.4\, (-0.76)$.
The sign of  $d_m$ and thus of $\Omega_{\rm P}$ in our model  results 
from
the mass of the lowest-lying strange scalar being greater than 
that of the lowest-lying isovector scalar.
As per our earlier classification, 
${\Omega_{\rm IB}^{(4)}}=
{\Omega_{\rm IB}^{(4),i}}
+ {\Omega_{\rm IB}^{(4),ii}}
+ {\Omega_{\rm IB}^{(4),iii}}
+ {\Omega_{\rm IB}^{(4),iv}}$, so that with 
$d_m=-2.4\,(-0.76)$, we have 
${\Omega_{\rm IB}^{(4),iv}} = -0.79 \,(-0.21)$. 
Estimating ${\Omega_{\rm IB}^{(4)\,,ii}}$ using 
the $\chi$-dependent $E_i^-$ yields
${\Omega_{\rm IB}^{(4),ii}} = -0.12\, (-0.03)$. 
${\Omega_{\rm IB}^{(4),iii}}$ has been partially
determined through the inclusion of 
$\pi^0$-$\eta^\prime$ mixing in 
$\Omega_{\eta+\eta^\prime}=0.25 \pm 0.05$~\cite{dght,buge}.
Using the result
${\Omega_{\rm IB}^{(2)}} + {\Omega_{\rm IB}^{(4),iii}} 
= 0.16 \pm 0.03$~\cite{ecker} and 
neglecting ${\Omega_{\rm IB}^{(4),i}}$, as the ill-known 
$E_i^-$ do not warrant such a calculation, 
we estimate, finally, 
that $\Omega_{\rm IB}=\Omega_{\rm IB}^{(2)} + \Omega_{\rm IB}^{(4)}
\sim -0.05 \ra -0.78$. 
For reference, 
note that $\Omega_{\rm IB}^{(2)}\sim 0.13$. 
The large value 
of $\Omega_{\rm IB}^{(4)}$ 
is driven by the numerical prefactor of 
Eqs.~(\ref{omegaP},\ref{omegaetap}) --- the contributions in  
$\Omega_{\rm IB}^{(4)}$ are ``naturally'' of the same size as 
$\Omega_{\rm IB}^{(2)}$. 
Thus we find a very large correction to the value of 
$\Omega_{\eta+\eta^\prime}=0.25\pm 0.05$, 
used in ``central value'' of \epsrat. 
The large negative change in 
$\Omega_{\rm IB}$ found in \opfour 
generates a substantial increase in \epsrat.

The $\Omega_{\rm IB}$ we calculate impacts \epsrat in a
significant manner. 
Our estimate of  $\Omega_{\rm IB}$   from 
the specific  $m_d\ne m_u$   effects we consider
ranges from  $-0.05 \ra -0.78$; this range exceeds the 
central value, $\Omega_{\eta+\eta^\prime}=0.25 \pm 0.05$, 
used in earlier analyses and 
reflects a variation in 
\epsrat of more than a factor of two.  
The presence of unknown low-energy constants implies that
we lack a reliable way to calculate the effects we consider. 
Such limitations, however, underscore the need for a 
larger uncertainty in the Standard Model prediction of \epsrat.

The collaboration of G. Valencia is gratefully acknowledged.
The work of S.G. 
is supported in part by the 
U.S. DOE under contract number DE-FG02-96ER40989.

\end{document}